\documentclass[traditabstract]{aa}

\usepackage{amsmath}
\usepackage{txfonts}
\usepackage{graphicx}
\usepackage{amssymb}
\usepackage{epstopdf}
\usepackage{natbib}
\usepackage[para]{threeparttable}

\bibpunct{(}{)}{;}{a}{}{,}
\begin{document}

\authorrunning{Shah et. al.}
\titlerunning{Improving GW parameter estimation for compact binaries
  with LISA}

\title{Using electromagnetic observations to aid gravitational-wave
  parameter estimation of compact binaries observed with LISA}

\author{
S.~Shah \inst{\ref{radboud},\ref{nikhef}} \and
M.~van~der~Sluys \inst{\ref{radboud},\ref{nikhef}} \and
G.~Nelemans \inst{\ref{radboud},\ref{nikhef},\ref{leuven}}  
}

\institute{
Department of Astrophysics/ IMAPP, Radboud University Nijmegen, P.O. Box 9010, 6500 GL Nijmegen, The Netherlands, \email{s.shah@astro.ru.nl} \label{radboud} \and
Nikhef – National Institute for Subatomic Physics,  Science Park 105,  1098 XG Amsterdam, The Netherlands \label{nikhef} \and
Institute for Astronomy, KU Leuven, Celestijnenlaan 200D, 3001 Leuven, Belgium \label{leuven} 
}

\date{Received / Accepted }

\abstract
{We present a first-stage study of the effect of using
  knowledge from electromagnetic
  (EM) observations in the gravitational wave (GW) data analysis of
  Galactic binaries that are predicted to be observed by the new
  \textit{Laser Interferometer Space Antenna} in the low-frequency range,
  $10^{-4} \mathrm{Hz}<f<1 \mathrm{Hz}$. In particular, we examine    
  the extent to which the accuracy of GW parameter estimation improves if we use
  available information from EM data. We do this by investigating  
  whether correlations exist between the GW parameters that describe
  these binaries and whether some of these parameters are also available from
  EM observations. We used verification binaries, which are
known as the guaranteed sources for \emph{eLISA} and will test the
functioning of the instrument. We find that of the seven parameters that
characterise such a binary, only a few are correlated. The most 
useful result is the strong correlation between amplitude and
  inclination, which can be used to constrain the parameter uncertainty
  in amplitude by making use of  the constraint of inclination from EM
  measurements. The improvement can be up to a factor of $\sim6.5$,
  but depends on the signal-to-noise ratio of the source data. Moreover, we find that this
  strong correlation depends on the inclination. For mildly face-on
  binaries ($\iota \lesssim 45^{\circ}$), EM data on inclination can improve the 
estimate of the GW amplitude by a significant factor.  However, for
edge-on binaries ($\iota \sim 90^{\circ}$), the inclination can be
determined accurately from GW data alone, thus GW data can be used
to select systems that will likely be eclipsing binaries for EM follow-up. }

\maketitle

\section{Introduction}
The space-based gravitational wave (GW) detector in consideration by
ESA, \textit{eLISA}, is expected to observe
millions of compact Galactic binaries \citep{2009CQGra..26i4030N, 2012arXiv1201.3621A}
with periods shorter than about a few hours, amongst other astrophysical
sources, and resolve several thousand of
these binaries \citep{2012arXiv1201.4613N}. About 50 
compact binary sources have been observed at optical, UV, and
X-ray wavelengths \citep[e.g.][]{2010ApJ...711L.138R}. The types of binaries
known to us are interacting systems (AM~CVn stars, ultra-compact X-ray
binaries, and cataclysmic variables) and detached systems (double white
dwarfs (WDs) and double neutron stars \citep{2009CQGra..26i4030N,
  Nelemans_wiki}).  The AM~CVn stars are binary systems where a WD
accretes matter from a low-mass, helium-rich (hydrogen-deficient) 
object \citep{2010PASP..122.1133S}. Their 
mass transfer is driven by GW radiation loss. The known ultra-compact
X-ray binaries consist of neutron stars that are accretors whose
donors are inferred to be either helium rich or carbon/oxygen rich
\citep{2005A&A...441L...1I}. Double WDs are predicted to be the most
common systems, which sometimes tend to be the outcome of many binary evolutionary
paths \citep{1984ApJ...277..355W}. Of all these known systems, a handful lie in
the \textit{eLISA} band and will be individually detected. These are
known as \textit{verification binaries} 
since they are guaranteed sources for the detector. Parameter
uncertainties in the verification binaries and Galactic binaries in
general have been studied in the literature extensively by using Fisher information
matrix (FIM) analyses \citep[e.g.][]{1998PhRvD..57.7089C,
  2002ApJ...575.1030T, 2006CQGra..23S.809S}. These studies have been
done for various configurations of classic \textit{LISA}
\citep{LISA-yellow_book}, which was designed to have a larger baseline
of five million km with six laser links 
interchanging between three stations located at the vertices of a
triangle that was to house two proof masses each \citep{2008CQGra..25f5005V}. Instead
\textit{eLISA} will have a baseline of one 
million km with four laser links interchanging between the proof
masses. The parameter uncertainties depend intricately on the
observation conditions and the geometry of the detector
\citep{2002ApJ...575.1030T} and the same is true for the (possible)
correlations between the parameters of a Galactic binary. In this
study, we wish to quantify whether any such correlations exist that
could be useful in constraining the GW parameter estimates. Some
of the GW parameters are the same as or related to 
electromagnetic (EM) parameters, (e.g. inclination$_{\mathrm{GW}}$
= inclination$_{\mathrm{EM}}$, $f_{\mathrm{GW}} = 2/P_{\mathrm{orb}}$,
etc. ). Thus, we can use an independent (EM) constraint of a (EM/GW) parameter
that correlates strongly with another (GW) parameter to improve the 
accuracy of the latter. This work provides a first step towards
developing strategic plans in observing those potentially
useful EM parameters that can improve the GW parameter
accuracy. In this paper, we present a FIM analysis to study  
  whether correlations within GW parameters exist. For the useful 
  correlations that we find, we predict quantitatively the improvement in
  the GW parameter when there is \textit{prior} EM data in the
  correlated parameter. The paper is structured in the following way. In
Section 2, we briefly summarise the signal models, both the instrumental and
foreground noises, and our data analysis. We present our
results and interpretations in Section 3. Finally, we discuss how the
results differ from the old \textit{LISA} detector in Section 4 and
present our conclusions in Section 5.  

\section{Signal modelling and data analysis}
\subsection{Gravitational wave signals from a Galactic binary}
We consider three verification sources, AM~CVn, SDSS~J0651+2844
(hereafter J0651), and RX~J0806.3+1527 (HM~Cnc), whose physical parameters are
summarised in Table~\ref{tab:verf_bin}.  

\begin{table*}
\centering
\begin{threeparttable}
\caption{Physical properties of verification binaries based on
  observations summarised in \cite{Nelemans_wiki}. The signal-to-noise
  ratio (S/N) 
  listed in the second column is from GW data analysis, as explained in
  Section 3.}
\label{tab:verf_bin}
\begin{tabular}{c c c c c c c c c c c }
\hline \hline
& $\mathrm{S/N} $& $m_{1}$[$M_{\odot}$] & $m_{2}$[$M_{\odot}$] &
$d$[kpc] &$\mathcal{A}^{*}[\times 10^{-22}]$&  $P_{\mathrm{orb}}$[s] & $\iota $[$^\circ$] & $\dot{P}_{\mathrm{orb}}$[s/s] & $  \beta$[rad] & $\lambda$[rad]  \\
\hline
J0651 &$10.7$ & $0.55$\tnote{a}  & $0.25$\tnote{a}   & $\sim 1$\tnote{a} & $ 1.67$&$765.4\pm7.9$\tnote{a} & $86.9^{+1.6}_{-1}$\tnote{a} & - & $0.101$ & $1.769$  \\
 AM~CVn &$ 11.5$ & $0.71$\tnote{b} & $0.13$\tnote{b}  & $0.606^{+0.135}_{-0.93}$\tnote{c} & $1.49$& $1028.73$\tnote{d} & $43\pm2$\tnote{c}  & - & $0.653$ & $2.974$  \\
HM~Cnc & $ 39.7$ & $0.55$\tnote{e} & $0.27$\tnote{e} & $5$\tnote{e} & $6.38$& $ 321.529$\tnote{f} & $\approx38$\tnote{e} & $3.75\times10^{-11}$\tnote{f} & $-0.082$ & $2.102$   \\
    \hline
      \end{tabular}
      \begin{tablenotes}
        \item[a] \cite{2011ApJ...737L..23B}  
        \item[b] \cite{2006MNRAS.371.1231R}   
        \item[c] \cite{2007ApJ...666.1174R}  
        \item[d] \cite{1999PASP..111.1281S}  
       \item[e] \cite{2010ApJ...711L.138R}
        \item[f] \cite{2005ApJ...627..920S}
        \item[*] Eq. \ref{eq:amplitude}
     \end{tablenotes}
\end{threeparttable}
\end{table*}

AM~CVn and HM~Cnc are mass-transferring systems known to astronomers 
as AM~CVn binary systems, which were described in the introduction. J0651
\citep{2011ApJ...737L..23B} is an eclipsing detached WD binary system
that was spectroscopically identified in the Sloan Digital Sky Survey (SDSS) 
catalogue. AM~CVn and  J0651 can be modelled as monochromatic GW
sources, which means that they are described by seven parameters:
dimensionless amplitude ($\mathcal{A}$), 
frequency ($f$), polarisation angle ($\psi$), initial GW phase
($\phi_{0}$), inclination ($\cos \iota$), ecliptic 
latitude ($\sin \beta$), and ecliptic longitude ($\lambda$). HM~Cnc can be
modelled as a mild chirper with an 
additional eighth parameter, the chirping frequency 
($\dot{f}$).  The two polarised gravitational waveforms used in the strain
for slowly evolving binaries are given by \citep[e.g.][]{2004PhRvD..70b2003K}
\begin{equation}
\label{eq:h_plus}
 h_+(t) = \mathcal{A} \; \frac{1+\cos^2\iota}{2} \,
\cos \left( 2\pi f t+\pi \dot{f}t^2 +\phi_0 \right) ; 
\end{equation}
\begin{equation}
\label{eq:h_cross}
 h_\times(t) = \mathcal{A} \; \cos\:\iota \:\;
\sin \left( 2\pi f t+\pi \dot{f}t^2 +\phi_0 \right), 
\end{equation}
where 
\begin{equation}
\label{eq:amplitude}
\mathcal{A} = \frac{4(\mathrm{G}\;\mathcal{M})^{5/3}}{c^4 \; d}
(\pi f)^{2/3}
\end{equation}
and 
\begin{equation}
\label{eq:fdot}
\dot{f} = \frac{96}{5} \frac{f}{\mathcal{M}} (\pi f
\mathcal{M})^{8/3}. 
\end{equation}
In these expressions, $\mathcal{M} \equiv
(m_1m_2)^{3/5}/(m_1+m_2)^{1/5}$ is the chirp mass and $d$ is the distance
to the source. The monochromatic waveforms are given by setting
$\dot{f} = 0$ in the expressions above.

\subsection{Detector response to the GW signals}
If a GW signal is present, then
the output of a detector will contain the strain, $h(t)$, and noise, $n(t)$. Thus,
the detector registers
\begin{equation}
s(t) = h(t;\vec{\theta}) +n(t), 
\end{equation}
where $\vec{\theta}$ is a vector characterising the seven (or eight)
parameters of the binary. Since \textit{eLISA} will have a 
motion around the Sun and a cartwheeling motion around its
centre of mass, a monochromatic signal from a WD binary will be
modulated in its amplitude, frequency, and phase in complicated
ways. The resulting signal will be spread over a range of frequency
bins of the detector \citep{2003CQGra..20S.163C}. Galactic
binaries typically radiate monochromatic signals at low frequencies\footnote{low
  $f$ is  a function of the detector transfer frequency, $f_*$,$ f \ll
  f_*$. The $f_*$ is defined according to the detector armlength, 
  $L$, i.e. $f_* \equiv c/(2\pi L)$. For \textit{eLISA},
  $L=10^9 $m, $f_* \approx 5\times 10^{-2}$Hz} and thus the response at
the detector can be written as \citep{2003CQGra..20S.163C}
\begin{equation}
h(t) = A(t) \: \mathrm{cos}\Psi(t), 
\end{equation}
where 
\begin{equation}
\label{eq:amplitude_mod}
 A(t) = [(F^+(t)h_+(t))^2 + (F^{\times}(t)h_{\times}(t))^2]^{1/2}. 
\end{equation}
The functions $F^{+, \times}(t)$ are the antenna beam patterns of the detector, and
they depend on the source's sky position ($\lambda$, $\beta$), its
orientation ($\psi$), and the detector
configuration. The phase of the signal is given by
\begin{equation}
\Psi(t) = 2\pi f t+\phi_0+\Phi_D(t)+\Phi_P(t), 
\end{equation}
where $\Phi_D(t), \Phi_P(t)$ are frequency (Doppler) and phase
modulations, respectively \citep{2003CQGra..20S.163C}.
Doppler modulation is given by 
\begin{equation}
\label{eq:doppler_mod}
\Phi_D(t) = 2 \pi f L/c \;\; \sin \beta \; \cos(2 \pi
f_m t- \lambda),  
\end{equation}
where $f_m = 1/$year is the modulation frequency.  The phase modulation is given by
\begin{equation}
\label{eq:phase_mod}
\Phi_P(t) = - \arctan \left ( \frac {F^{\times} \;  h_{\times}} {F^+ \; h_+}\right). 
\end{equation}
Different architectures of the triangular 
space-based interferometer generate a number of independent
data streams that provide different responses to the incoming GW signal
\citep{2008CQGra..25f5005V}. For the most recent 
interferometer design in consideration, \textit{eLISA}, the output 
is a single unequal-arm Michelson data stream, \textit{X}. 
This is a linear combination of phase shifts 
measured at the different spacecraft (by comparing the incoming light with a 
local reference source) shifted in time in such a way as to represent 
interference between two light beams travelling through the arms of the 
detector in opposite ways, which is a particular implementation of so-called \textit{time 
delay interferometry} (TDI, \citet{1999ApJ...527..814A}. \citet{2005PhRvD..72d2003V} provide a
detailed description of how TDI works and an explanation of why it
produces an interferometry signal  in which the phase shift induced by
a passing GW is preserved, while the much  
larger shifts induced by instrumental noise are strongly suppressed. We made
 use of the existing numerical software \textit{Synthetic LISA}
 \citep{2005PhRvD..71b2001V} to simulate accurate time-domain series
 of the instrumental noise and the GW signals in the form of these TDI
 observables. The data stream is a discrete series for a given
 observational time $T_{\mathrm{obs}}$, where the samples 
are separated by $\Delta t$. The detector response to GWs and instrumental
noises have been discussed extensively in the literature
\citep [e.g.][]{1998PhRvD..57.7089C, 2003PhRvD..67b2001C,
2004PhRvD..70b2003K,  2005PhRvD..71b2001V}, hence we only summarise 
the most essential expressions relevant to our data analysis. 
In \textit{Synthetic LISA}, the strain at the detector, $s(t)$, is modelled as the
TDI $X$ observable. This is the quantity we work with in our data
analysis. 

\subsection{Noise}
There are two types of noise to consider: instrumental
noise and Galactic foreground noise due to unresolved compact binaries. For the
particular geometry of \textit{eLISA} used here, the
instrumental noises (mostly from shot noise), acceleration noise, and
other types of noise (e.g optical bench noise), are characterised by their
power spectral densities (PSDs\footnote{Note there is no PSD for laser
noise, since we assume that it is completely cancelled in the TDI
observable $X$.}) of $2.31\times 10^{-38} f^2$, $6\times
10^{-48}f^{-2}$, and $2.76\times 10^{-38}f^2$, respectively, in units of
Hz$^{-1}$.  The instrumental noise is modelled as a random, Gaussian
process. We note that the sampling time, $\Delta t$, should
be carefully chosen while simulating the instrumental
noise in order to correctly interpret the TDI observables. The choice
of $\Delta t$ should correspond to a frequency that is several times
higher than the highest frequency where the TDI responses have to be
analysed \citep{2005PhRvD..71b2001V}. This means that since AM~CVn and
J0651 have relatively low frequencies they can be 
analysed with samples of $\Delta t = 64$s, whereas for HM~Cnc, we
need a lower sampling time of at least $\Delta t = 16$s (see Figure
~\ref{fig:inst_noises} in Appendix B). 

The foreground noise from the Galactic binaries is simulated
using the \textit{Lisasolve} \citep{lisasolve} software where every binary
(monochromatic and/or mild chirper\footnote{Mild implies that $\dot{f} /f
  \ll 1/ \mathrm{T_{obs}}$, where $\mathrm{T_{obs}}$ is the observational time.}) is modelled in
the frequency domain \citep{2007PhRvD..76h3006C}. This differs from
simulating signals using \textit{Synthetic LISA}, where the signal is
modelled accurately in the time-domain.  
\textit{Lisasolve} instead makes use of the very slowly evolving nature of the
binaries to approximately model the signals both directly and speedily in
the frequency domain. We only use double-detached WD binaries because they 
form the majority of the foreground noise \citep{2001A&A...375..890N,
  C3lisa}. We include more than $2.7 \times 10^7$ detached
Galactic double WDs from a simulation with the same assumptions  
about binary evolution and Galactic distribution as those in
\citet{2004MNRAS.349..181N}, but with (about a factor of ten) higher intrinsic
resolution. These  
assumptions have also been used for the simulations of the Galactic binaries 
on which the Mock LISA Data Challenge (MLDC) rounds are based
\citep{2011PhRvD..84f3009L}. 

The \textit{unresolvable} Galactic foreground noise is obtained by
iteratively subtracting the \textit{resolvable} sources (i.e. binaries
expected to be detected) from the simulated population as follows: 
\begin{enumerate}
\item The S/N (see Eq. \ref{eq:snr}) is computed for each
    binary against the Gaussian instrumental noise and
    \textit{initial} Galactic foreground noise for $\mathrm{T_{obs}}$
    of two years. The initial foreground noise is calculated using the initial
    catalogue of $2.7 \times 10^7$ detached Galactic double WDs. 
\item All the sources with S/N $>5$ are removed from the initial  
catalogue/dataset. The \textit{reduced} dataset is used to simulate the
reduced Galactic foreground.
\item Using the reduced Galactic foreground and the same
    instrumental noise, the S/N for each of the binaries in the reduced
    catalogue is calculated. The process is iterated with step 2.
\end{enumerate}

We applied a perfect subtraction of the bright
  sources where any spurious effects in the data set were not taken into
  account (since the sources were removed from the population before
  generating the signals). A Markov chain Monte Carlo based data
  analysis of the MLDC shows that all the recovered parameters of the subtracted 
  sources have a strong peak of zero bias when compared to their
  injected values in the training data set
  \citep{2011PhRvD..84f3009L}. Thus, this perfect subtraction scheme is not
  expected to introduce strong biases in our results. From our subtraction procedure 
  outlined above, we estimate the number of 
resolved WD binaries to be $\sim 11,000$. Using
  the same subtraction procedure and S/N threshold,
  \cite{2012arXiv1201.4613N} estimate the number of bright sources to
  be half of our estimate, although their estimate is only for
a $\mathrm{T_{obs}}$ of one year. Furthermore, using a higher S/N
  threshold of seven, the number of resolvable sources is 3,000 for
  an observation time of two years \citep{2012arXiv1201.3621A}. If we
  use a threshold S/N of seven, we find 4,500
resolvable sources. The PSD of
the unresolvable Galactic background is consistent with 
findings in the literature where the foreground noise for \emph{eLISA}
is almost at the level of instrumental noise for this detector
\citep{2012arXiv1201.4613N}, unlike in the case of the classic
\emph{LISA} where the foreground noise was predicted to dominate at $f
\lesssim 3$mHz \citep{2001A&A...375..890N, PhysRevD.73.122001}.    

\subsection{Data analysis}
For GW sources with known waveforms, one can use
matched filtering methods \citep{1992PhRvD..46.5236F, 1994PhRvD..49.2658C}
to extract the signal parameters and estimate their
uncertainties. Consequently, when the noise is Gaussian, the
parameter uncertainties are given by their joint Gaussian probability
distribution function \citep{1998PhRvD..57.7089C} 
\begin {equation}
p(\sigma_{\vec{\theta}}) = \sqrt{\mathrm{det(\Gamma/2\pi)}} \; \;
\mathrm{exp}\left(-\frac{1}{2} \; \Gamma_{ij} \; \sigma_{\theta_i} \;
  \sigma_{\theta_j}\right), 
\end{equation}
where $\Gamma$ is known as the Fisher information matrix (FIM) given
by 
\begin {equation}
\label{eq:fisher}
\Gamma_{ij} \equiv \left( \frac{\partial h}{\partial
    \theta_i} \;  \middle | \; \frac{\partial h}{\partial \theta_j} \right). 
\end{equation}
The inner product $(...|...)$ is a generalisation of the
time-domain correlation product and is conventionally defined as
\begin {equation}
\label{eq:scalar_product}
(a|b) = 4 \int_{0}^{\infty} df \; \frac{\tilde{a}^{*}(f) \;
  \tilde{b}(f)} {S_n(f)} \simeq \frac{2}{S_n(f_0)} \int_0^{T_{\mathrm{obs}}} dt \;\: 
a(t) \; b(t).
\end{equation}
Eq.\,\ref{eq:scalar_product}\footnote{The latter equality follows from
Parseval's theorem.} holds for quasi-monochromatic binaries
that have an almost constant noise PSD, $S_n(f)$, in the frequency
region where the binary radiates
\citep{1998PhRvD..57.7089C}. The S/N of a source is defined
  as the inner product of a signal with itself 
\begin {equation}
\label{eq:snr}
\mathrm{S/N}^2 = (h|h).
\end{equation}
In the limit of signals with a high signal-to-noise ratio (S/N $\gg 1$), the inverse of the
FIM gives the variance-covariance matrix $\mathcal{C} =
\Gamma^{-1}$. The diagonal elements  $\mathcal{C}_{ii}$
give variances (or mean square errors) in each
parameter, $\langle (\sigma_i)^2 \rangle$, and the off-diagonal
elements describe the covariances (or correlations) 
between them. For each of our verification binaries, we calculate this
matrix to investigate the correlations between the binary
parameters. The derivative in Eq.\,\ref{eq:fisher}, $\partial
h/\partial \theta_i$, is numerically calculated in the time-domain 
\begin {equation}
\label{eq:derivative}
h' (t; \theta_i) \equiv \frac{h(t;
    \theta_i+d\theta_i) - h(t; \theta_i-d\theta_i)}{2 d\theta_i},
\end{equation}
where $d\theta_i$ should be chosen carefully. In general, the quantity $d\theta_i$
should be as small as the machine accuracy allows for, but not too
large to suffer from the truncation error\footnote{This error comes
  from higher-order terms in the Taylor-series expansion, $h(x+dx) = h(x) +
  dx\; h'(x) + \frac{1}{2}\;dx^2\;h''(x) + ... $}. Thus, for well-behaved 
functions, $d\theta_i \sim \sqrt{\epsilon} \: \theta_c$, where
$\epsilon \sim 10^{-16}$ is the machine accuracy and $\theta_c$ is
some typical value of the corresponding parameter \citep{2002nrc..book.....P}.
To find a good choice of $d\theta_i$, we compute $\sigma_i$ for a range
of $d \theta_i$ with logarithmic intervals and select the value for
each corresponding parameter around which the standard deviations
become stable. By stable, we mean that increasing or decreasing
$d\theta_i$ by an order of magnitude should lead to values of $\sigma_i$ that vary by
no greater (smaller) factor than 1.1 (0.9). An example of
the stabilisation of the
variance-covariance matrix is provided in Table~\ref {tab:stabilise} of 
Appendix A for the case of AM~CVn. We perform this stability check for all
our verification binaries. For instance, in the case of AM~CVn, $d\theta_{\mathcal{A}} \sim 
10^{-30}$, $d\theta_{f} \sim 10^{-11}$, $d\theta_{\iota, \psi}
\sim 10^{-8}$ etc. All our analysis is done for $T_{\mathrm{obs}} =  2$ years.  

\section{Results}
Here we list the variance-covariance matrix $\mathcal{C}$ for
AM~CVn,  HM~Cnc, and J0651 and 
discuss the strongest correlations that we found. The
off-diagonal elements are specified by the normalised correlations,
$\mathrm{c}_{ij}$,  and the diagonal elements are quoted as the
square root of the variances (or standard deviations),
$\mathrm{c}_{ii}$, i.e.
\begin{equation}
\mathrm{c}_{ij} =
\frac{\mathcal{C}_{ij}}{{\sqrt{\mathcal{C}_{ii}\mathcal{C}_{jj}}}},
\;\;\; \mathrm{c}_{ii} = \sqrt{\mathcal{C}_{ii}} \equiv \sigma_{i}.
\end{equation}
Thus, $\mathrm{c}_{ij}$ can have values in the range $[-1, +1]$ where,
$\mathrm{c}_{ij} = +1$ means maximally correlated and $\mathrm{c}_{ij}
= -1$ means maximally anti-correlated. We consider highly correlated
parameters $\theta_i, \theta_j$ to be those for which $|\mathrm{c}_{ij}| > 0.8$. 
In the later subsections, we show that these correlations are affected when
the inclination is varied. 

The correlation matrices for the three verification binaries with
observed parameters as shown in Table~\ref{tab:verf_bin} are listed
below. The measured GW parameter values are  shown above each matrix. Since
there are no EM measurements for $\phi_0$ and $\psi$, we set them to
$\pi$ and $\pi/2$, respectively, for all three binaries. The choice of
the $\phi_0$  value does not influence the results shown below,
whereas the choice for the $\psi$ has an effect that will be addressed
later. In the matrices, we take the medians of values obtained using 50
different instrumental-noise realisations and the standard deviations
about these medians ($\sigma$(S/N) and $\sigma(\sigma_i)$). They are
listed above and below each of the corresponding matrices, 
respectively. The values of S/N we find here
are comparable to the ones found by other groups \citep{C3lisa}. The
normalised correlations are accurate within the quoted precision and
hence their standard deviations from the instrumental noise are not listed.  \\\\
J0651, S/N $= 10.72\pm 0.23$.\\
Strong correlations (as defined above) are printed in bold face.
\resizebox{\columnwidth}{!}{
\bordermatrix{~ & \mathcal{A} & \phi_0 & \mathrm{cos}\iota & f & \psi & \mathrm{sin}\beta & \lambda  \cr
                 \theta_i&1.670\times10^{-22} & \pi &  0.01& 2.614\times10^{-3}&  \pi/2&0.10&1.77 \cr \hline \cr
                   \mathcal{A}  & 1.564\times10^{-23}& 0.01& -0.05& -0.02& \;\;\;0.02& \;\;\;0.03& -0.08\cr
                  \phi_0 & & 0.208& -0.01& \mathbf{-0.89}& -0.02& \;\;\;0.13& -0.13\cr
                 \mathrm{cos}\iota & && 0.043& \;\;\;0.01& \;\;\;0.02& -0.06& \;\;\;0.34\cr
                  f & &&& 8.375\times10^{-10}&\;\;\;0.01& -0.17& \;\;\;0.16 \cr
                  \psi & &&&&0.040& -0.03& \;\;\;0.09 \cr
                  \mathrm{sin}\beta  & &&&&&  0.069& \;\;\;0.09 \cr
                  \lambda  &&&&&&&  \;\;\;0.020  \cr \hline \cr
                  \sigma(\sigma_i)3.31\times10^{-25}&  0.004&0.001& 1.771\times10^{-11}& 0.001&0.001&0.000\cr}
}
\\\\\\
AM~CVn, S/N $=11.54 \pm 0.19$.\\
\resizebox{\linewidth}{!}
{
\bordermatrix{~ & \mathcal{A} & \phi_0 & \cos\iota & f & \psi &  \sin\beta & \lambda   \cr
                  \theta_i&1.494\times10^{-22} & \pi & 0.73& 1.944\times10^{-3}&  \pi/2&0.61&2.97 \cr \hline \cr
                  \mathcal{A}  & 1.084\times10^{-22} & \;\;\;0.29& \mathbf{-0.99}& -0.06& -0.30& -0.03& -0.60\cr
                  \phi_0 & &2.333 &-0.27& -0.03& \mathbf{-0.99}& \;\;\;0.26& -0.44 \cr
                  \cos\iota & &&0.580& \;\;\;0.06& \;\;\;0.28& \;\;\;0.04& \;\;\;0.60 \cr
                  f & &&&6.807\times10^{-10}& -0.03& -0.11& -0.19 \cr
                  \psi & &&&&1.170& -0.27& \;\;\;0.03 \cr
                  \sin\beta  & &&&&&0.029& \;\;\;0.03  \cr
                  \lambda&&&&&&&0.040\cr \hline \cr
                  \sigma(\sigma_i) &1.821\times10^{-24} &0.039&0.010&1.145\times10^{-11}&0.020& 0.000&  0.001\cr}
}
\\\\\\
HM~Cnc, S/N $= 39.75 \pm 0.82$.\\
\resizebox{\linewidth}{!}{
\bordermatrix{~ & \mathcal{A} & \phi_0 & \cos\iota & f & \psi & \sin\beta & \lambda \cr
                 \theta_i&6.378\times10^{-23} & \pi &  \;\;\;0.79& 6.22\times10^{-3}&  \pi/2&-0.08& \;\;\;2.10 \cr \hline \cr
                  \mathcal{A}  &1.236\times10^{-23}&  \;\;\;0.01& \mathbf{-0.99}& 0.00& -0.01& -0.12&  \;\;\;0.07\cr
                 \phi_0 &&0.916& -0.01& -0.14& \mathbf{-0.998}& -0.16& -0.08\cr
                 \cos\iota &&&0.169&  \;\;\;0.00&  \;\;\;0.01& -0.12& -0.07\cr
                  f&&&&2.257\times10^{-10}&  \;\;\;0.09&  \;\;\;0.29& -0.06\cr
                 \psi &&&&&0.455&  \;\;\;0.14& -0.06\cr
                  \sin\beta&&&&&&0.018& -0.06\cr
                 \lambda&&&&&&&0.002\cr \hline \cr
                  \sigma(\sigma_i)&2.5\times10^{-25}&  0.019&0.003& 4.6\times10^{-12}& 0.009&0.000&0.000\cr}
}
\\\\\\
HM~Cnc, with $\dot{f}$, S/N $=39.89 \pm 0.85$.  \\
\resizebox{\linewidth}{!}
{
\bordermatrix{~ & \mathcal{A} & \phi_0 & \cos\iota & f & \dot{f}&\psi & \sin\beta & \lambda \cr
                 \theta_i&6.378\times10^{-23} & \;\;\;\pi & 0.79& 6.22\times10^{-3}& -7.25\times10^{-16}&  \pi/2&-0.08&2.10\cr \hline \cr
                   \mathcal{A}  &1.240\times10^{-23}& \;\;\;0.0& \mathbf{-0.99}& \;\;\;0.02& -0.02& \;\;\;0.00& \;\;\;0.15& -0.08\cr
                  \phi_0& & \;\;\;0.907& \;\;\;0.00& -0.14& \;\;\;0.11& \;\;\;\mathbf{0.995}& -0.02& -0.04\cr
                  \cos\iota && &\;\;\;0.172& -0.02& \;\;\;0.02& -0.01& -0.15& \;\;\;0.08\cr
                   f & &&& 9.58\times10^{-10}& \mathbf{-0.97}& -0.05& -0.05& -0.28\cr
                  \dot{f}& &&&& 2.971\times10^{-17}& \;\;\;0.04& \;\;\;0.14& \;\;\;0.28\cr
                  \psi & &&&&&\;\;\;0.448& -0.01& -0.07\cr
                  \sin\beta&&&&&& & \;\;\;0.018& \;\;\;0.09\cr
                  \lambda&&&&&&&&  \;\;\;0.002 \cr \hline \cr
                  \sigma(\sigma_i)&2.6\times10^{-25}&  0.019&0.004&2.1\times10^{-11} &5.96\times10^{-19}  & 0.009&0.000&0.000&\cr}
}
\begin{figure*}
\centering
\includegraphics[width=17cm]{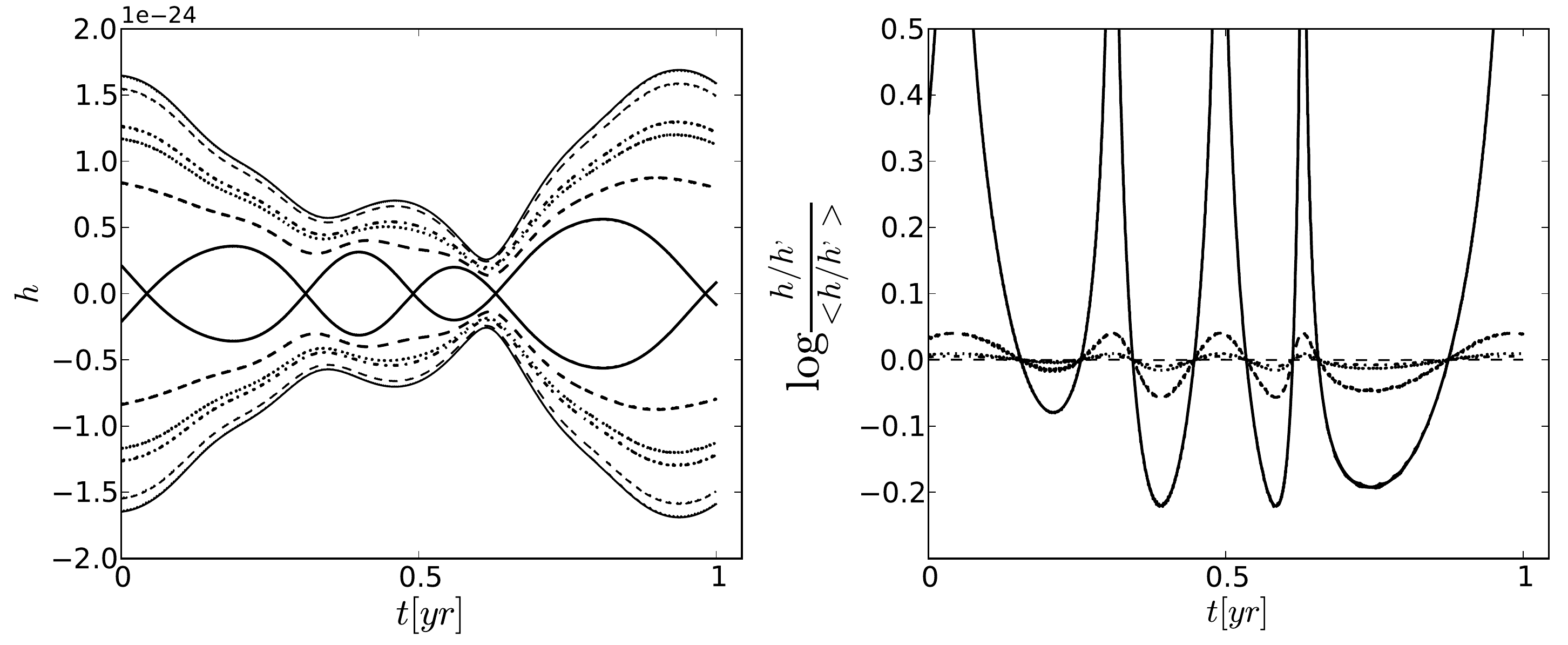}
\caption{Left: Envelopes of the GW signal (TDI $X$ observable) for binaries
  with parameters of AM~CVn, but for various inclinations as a
  function of time (innermost to outermost envelopes correspond to $\iota = 90^{\circ},
  60^{\circ}, 45^{\circ}, 40^{\circ}, 20^{\circ}, 0^{\circ}$
  respectively). The modulation of the signal is due
  to the complicated annual motion of \emph{eLISA} around the
  Sun. For low inclinations ($\iota = 0^{\circ}-45^{\circ}$), the
  signals only differ in amplitude, while the higher inclinations also
  differ in structure. To highlight this, we plot in the right panel
  the normalised ratio of the upper envelopes compared to the $\iota =
  0^{\circ}$ envelope using the same line styles (but now the $\iota =
  90^{\circ}$ is the outermost line). High inclination systems clearly
  have a unique structure, while low inclination envelopes only differ
  in amplitude, illustrating the degeneracy between $\iota$ and
  $\mathcal{A}$. } 
\label{fig:AMCVn_vs_incs} 
\end{figure*}

\subsection{The correlation between $\mathcal{A}$ and $ \cos \iota$} 
 \begin{figure*}
\centering 
\includegraphics[width=17cm]{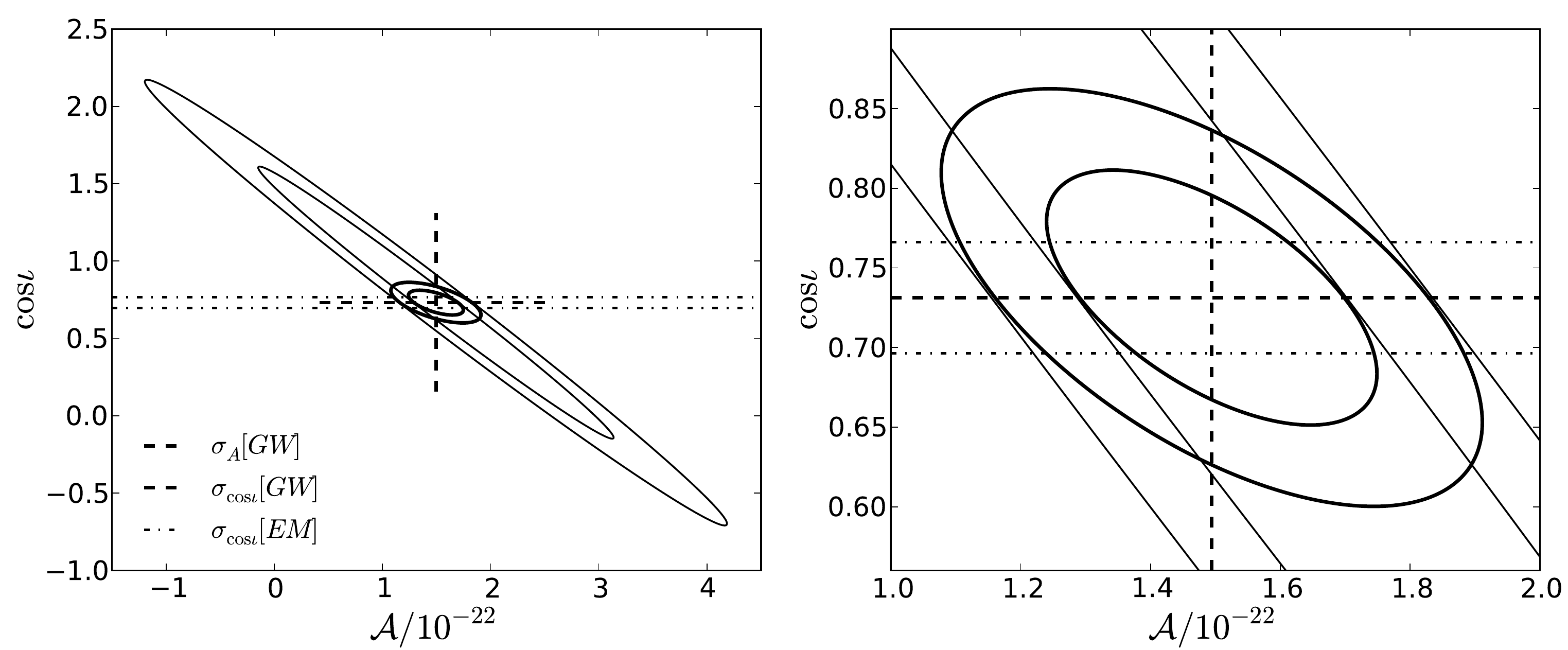}
\caption{ Left: Two-dimensional error ellipses of $\mathcal{A}$ and $\cos \iota$
  extracted from the variance-covariance matrix, $\mathcal{C}$, for
  AM~CVn. The two thin black lines represent 1-$\sigma$ and
  2-$\sigma$ ellipses and the black dashed lines represent
  the 1-$\sigma$ GW errors in $\mathcal{A}$ and $\cos \iota$ with
  the cross at the true value of the parameters. The unphysical values
  (i.e. negative numbers along the amplitude axis) are caused by the
  Gaussian tails of the parameter uncertainty about their true
  values. The right panel is a zoom of the area that is constrained by
  the 1-$\sigma$ EM error of $4^{\circ}$ shown in dash-dotted
    lines. The 1-$\sigma$ GW standard deviation in $\mathcal{A}$
  decreases from $\sim 1.08 \times 10^{-22}$ to $\sim 0.165 \times
  10^{-22}$, roughly a factor of 6.5. The corresponding error ellipses
  for the reduced PDF are shown as thick black ellipses for
  1-$\sigma$ and  2-$\sigma$. } 
\label{fig:AMCVn_Ai}
\end{figure*}
The amplitude and inclination of AM~CVn and HM~Cnc (when modelled
without the chirp, $\dot{f}$) are highly anti-correlated, with a value
of  $c_{\mathcal{A}\: \cos \iota} \approx - 0.99$. The reason for this
is that binaries with inclinations in the range from $0^{\circ}$ to
$\sim 45^{\circ}$ (face-on or close to face-on) have signals with the
same structure (see Figure~\ref{fig:AMCVn_vs_incs}), so the signals in this
range can be transformed into each other by simply changing either
the amplitude or the inclination. Thus, if we change the amplitude of one of
these signals within its GW uncertainty, we can obtain a very 
similar signal by changing its inclination within its GW 
uncertainty as well. However, for inclinations in the range
$\sim 45^{\circ} - 90^{\circ}$ (either close to edge-on or edge-on), this scaling does not
apply. This is because the signals in this range not only vary in
terms of the overall scale but also in their structure, when their inclination is
changed. This is especially evident for a binary with $\iota \sim
90^{\circ}$ (lines of $\iota = 60^{\circ}, 90^{\circ}$ in
Figure~\ref{fig:AMCVn_vs_incs}).  Since the inclinations of both
AM~CVn and HM~Cnc are within the range where similar signals can be obtained by
varying either the inclination or the strain amplitude, the parameters 
$\cos \iota$ and $\mathcal{A}$ are highly correlated. Furthermore since J0561 is an
edge-on system with $\iota \sim 90^{\circ}$, both $\cos \iota$ and 
$\mathcal{A}$ are distinguishable, as can be seen in
Figure~\ref{fig:AMCVn_vs_incs}.  This is potentially the  most useful
correlation from the GW analysis where an EM measurement of the
inclination of a binary can constrain the error in the strain
amplitude $\mathcal{A}$. For example for AM~CVn, an EM constraint in
$\iota$ with a 1-$\sigma$ accuracy of $4^{\circ}$ \citep{2006MNRAS.371.1231R}
improves the uncertainty in $\mathcal{A}$ from $1.084\times10^{-22}$ to
$1.65\times10^{-23}$ as shown in Figure~\ref{fig:AMCVn_Ai}. The
corresponding 1-$\sigma$ accuracy in $\mathcal{A}$ is estimated by selecting
the two-dimensional probability distribution function (2D PDF) points
that lie in the range $41^{\circ} < \iota < 45^{\circ}$. This improves
amplitude accuracy by a factor of 6.5 compared to the use of GW data alone.

\begin{figure}[!h]
\centering
\includegraphics[width=\columnwidth]{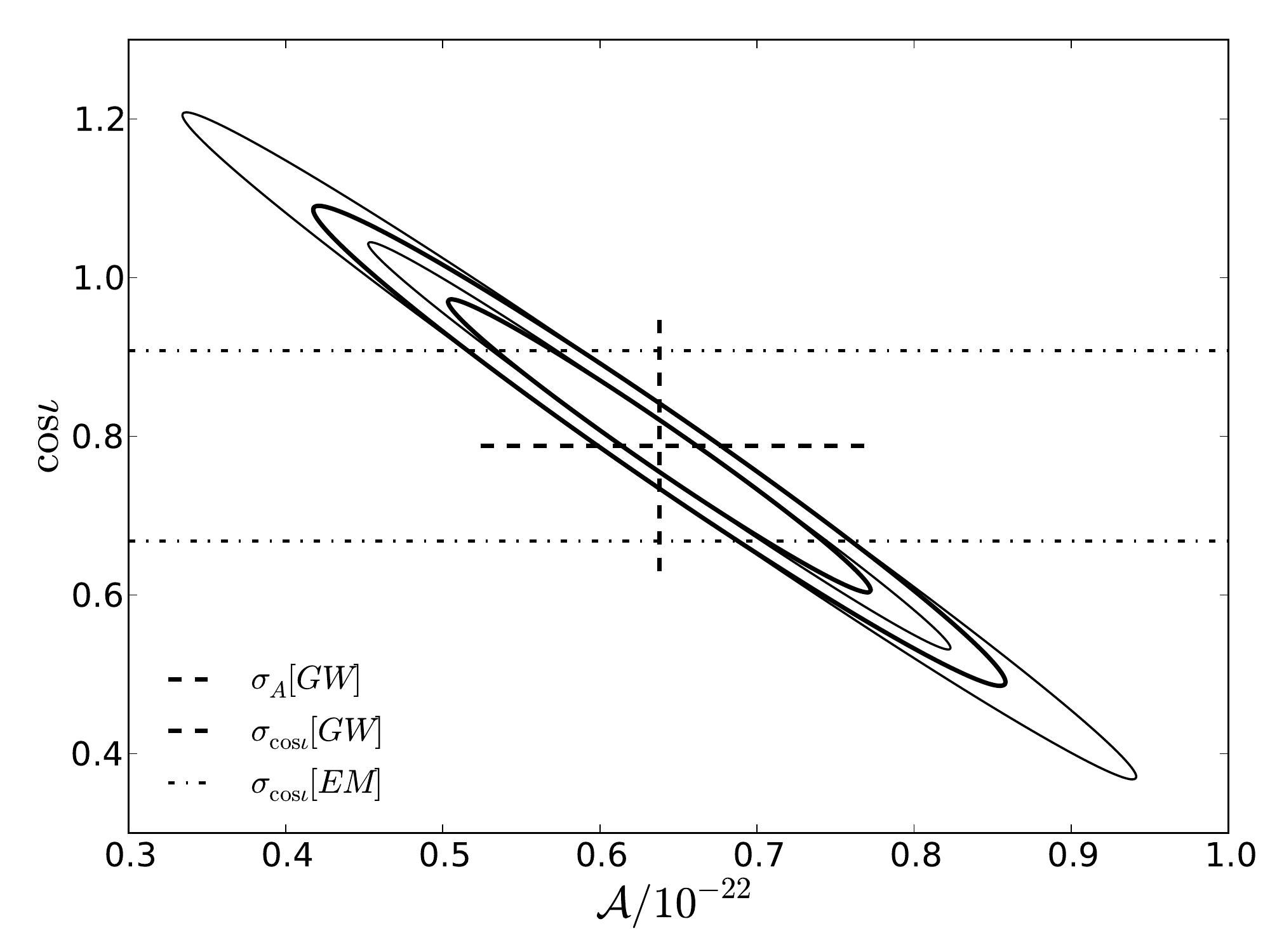}
\caption{\label{fig:HMCnc_Ai} Two-dimensional error ellipses  
  $\mathcal{A}$ and $\cos \iota$ extracted from the variance-covariance
  matrix, $\mathcal{C}$, for HM~Cnc. See Figure~\ref{fig:AMCVn_Ai} for
  details. The 1-$\sigma$ EM constraint of $7^{\circ}$ in inclination
  reduces the corresponding 1-$\sigma$ uncertainty in $\mathcal{A}$ from $\sim 1.22
  \times 10^{-23}$ to $\sim 0.89 \times 10^{-23}$, roughly a factor of 1.4.}
\end{figure}

HM~Cnc's inclination measurement derived from the EM observation is quite
uncertain compared to 
that for AM~CVn. This is mainly due to the large uncertainty in the mass ratio
and the assumption about the GW radiation that goes into estimating the
inclination \citep{2010ApJ...711L.138R}. Hence, if we take an EM constraint in
$\iota$ with a 1-$\sigma$ uncertainty of $7^{\circ}$, the amplitude 
improves by a factor of 1.34 as shown
in Figure~\ref{fig:HMCnc_Ai}. However, if we were to obtain a better 
$\iota$ constraint of $4^{\circ}$ for HM~Cnc, $\mathcal{A}$ would
improve by a factor of three, which is still a  
smaller improvement than that of the AM~CVn's. This is due to the
higher S/N of HM~Cnc: if we place AM~CVn at a closer distance such that
it has a higher SNR of $\sim 40$, the 1-$\sigma$ accuracy of
$4^{\circ}$ constrains the amplitude more tightly by a factor of 2.95, which is very
similar to the case of HM~Cnc. By constraining the strain amplitude,
we constrain the chirp mass, the distance, or a combination of the two
$\frac{\mathcal{M}^{5/3}}{d}$ (see Eq.\,\ref{eq:amplitude}). The GW
frequency, the third contributor to the amplitude, is already
determined very precisely. Additionally, if $\dot{f}$ is measurable
for a chirping binary, then from Eqs.\,\ref{eq:amplitude} and
\ref{eq:fdot} one can constrain the error in the distance
\citep{1996CQGra..13A.219S}.   

\subsection{Influence of the inclination, $\iota$}
\begin{figure*}
\centering
\includegraphics[width=17cm]{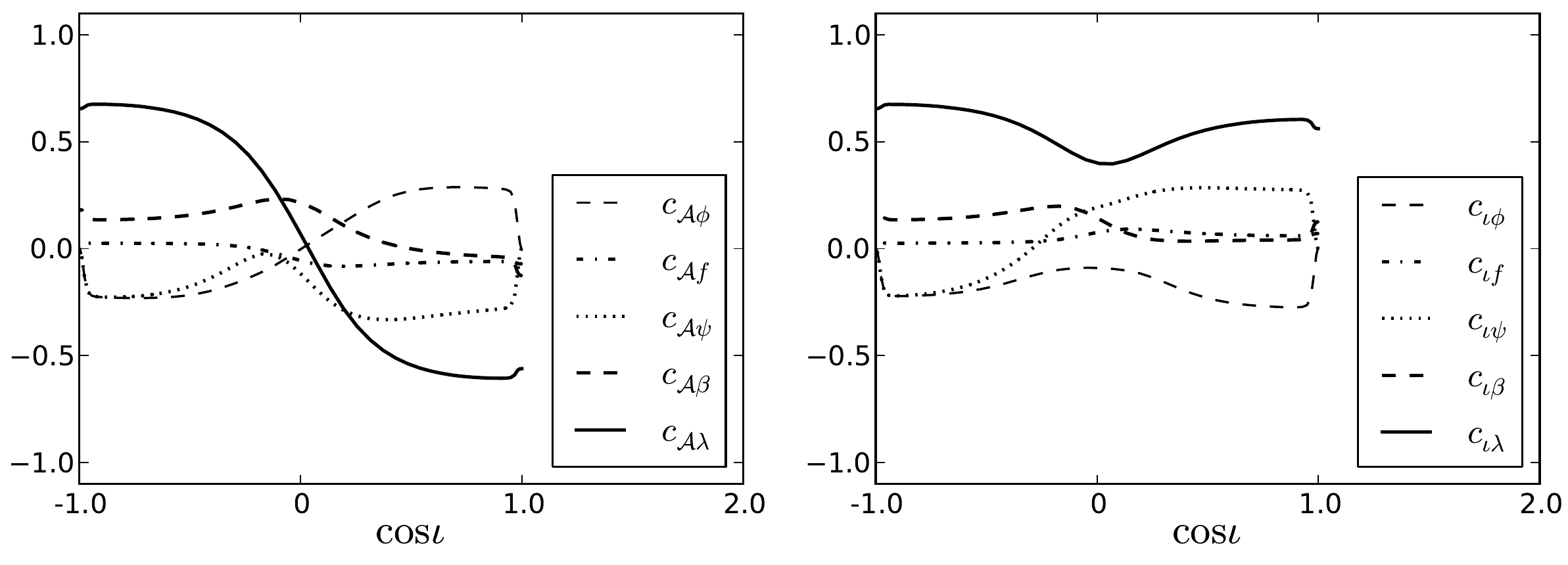}
\caption{Correlations of the remaining parameters with $\mathcal{A}$
  and $\iota$ for a binary with AM~CVn's parameter values as a
    function of $\cos \iota$. Most correlations are weak for all
    inclinations except $c_{\mathcal{A\lambda}}$.}
\label{fig:AMCVn_corrs_func_i}
\end{figure*}

We have seen above that $c_{\mathcal{A} \:\cos\:\iota}$ depends on the
inclination of the binary. For the special case of exactly face-on
binaries ($\iota = 0, \pi$), the Fisher matrix is ill-defined owing to 
the symmetry in the signal, namely, $h(t, \iota+d\iota) = h(t,
\iota-d\iota) $. The derivatives are zero, so the terms including $\iota$ in the
Fisher matrix are also zero, resulting in an ill-defined matrix that cannot be
inverted. In general, it is hard to track the behaviour of the
variance-covariance matrix from the FIM alone. Hence, to
determine the influence of inclination, we fix all parameters
except $\mathcal{A}$, $\iota$, and $\lambda$, and calculate the matrices $\Gamma$
and $\mathcal{C}$ for AM~CVn as a function of $\iota$. This can be
justified, as most correlations with $\mathcal{A}$ and $\iota$ are not
very strong; the largest correlation (of $\sim 0.6$) is for $\lambda$ as shown in
Figure~\ref{fig:AMCVn_corrs_func_i}. 
\begin{figure}[!h]
\centering
\includegraphics[width=\columnwidth]{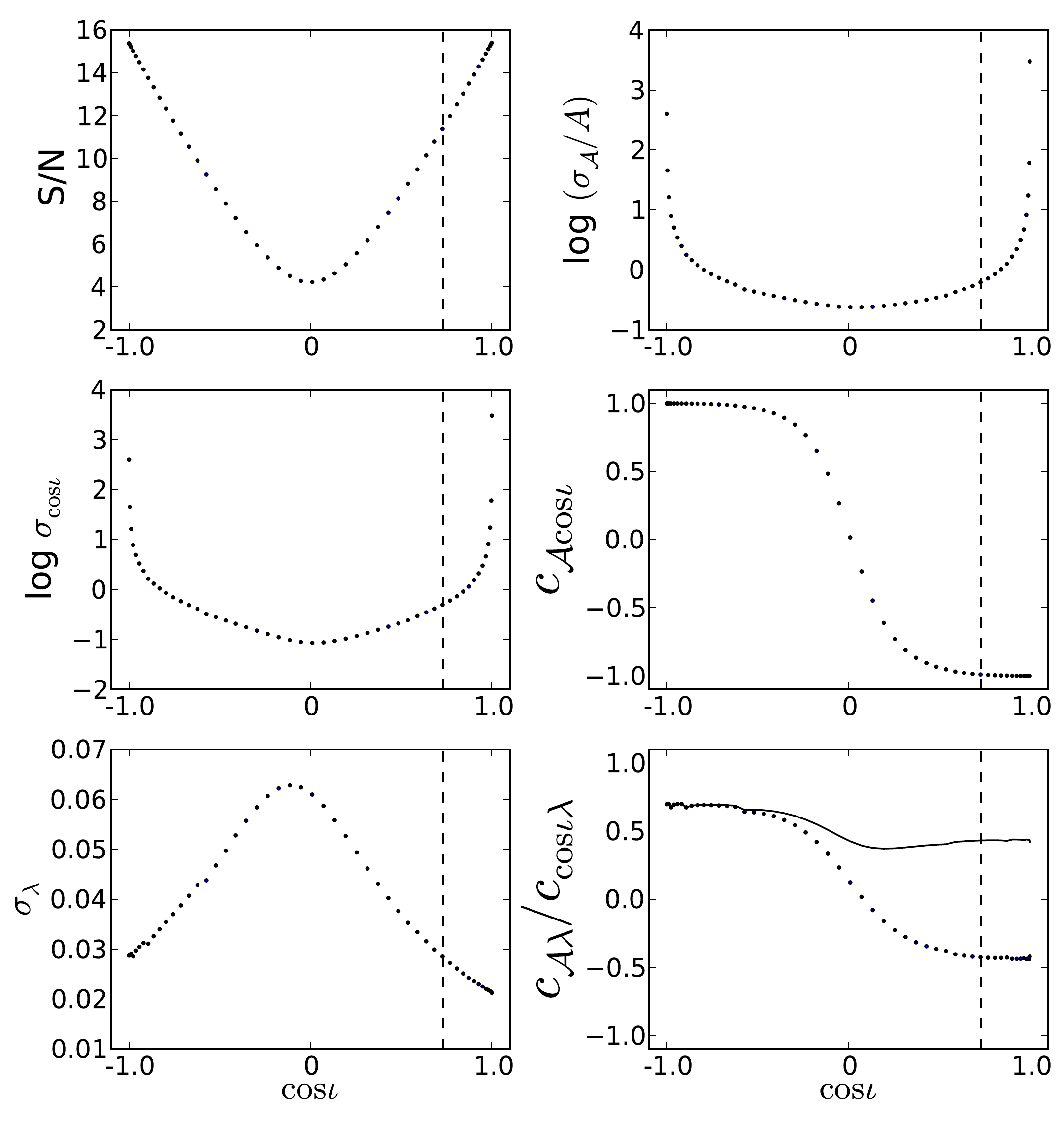}
\caption{ Signal-to-noise ratio, $\sigma_{\mathcal{A}}$, $\sigma_{\cos\iota}$ , $\sigma_{\lambda}$,
  $c_{\mathcal{A}\cos\iota}$, $c_{\mathcal{A}\lambda}$, and $c_{\cos\iota\lambda}$ as a
    function of $\cos \iota$, keeping  the other parameters fixed to
    the values of AM~CVn. The vertical dashed line is the measured value
    of $\iota$ for this binary, $43^{\circ}$. The best known 
    value for the amplitude is $\mathcal{A} = 1.49\times10^{-22}$. As
    expected, the S/N is higher for face-on than edge-on
    orientations. The relative uncertainties in the amplitude and
    inclination are very large for inclinations up to
    $10^{\circ}$. The bottom-right panel shows the normalised correlations
    $c_{\mathcal{A}\lambda}$ as the dotted line and
    $c_{\cos\iota\lambda}$ as the solid line.}
\label{fig:AMCVn_errs_func_i_3X3}
\end{figure}
This means that for all inclinations \[ \mathcal{C} \sim  \bordermatrix{~ & & \cr
                               & \mathcal{C}_{\mathcal{A} \:\mathrm{cos}\iota  \:\lambda} &  0 \cr
                               &  0 & \mathcal{C}_{\phi_0 \:f\: \psi\: \mathrm{sin}\beta }  \cr} \]
thus we can focus on $\mathcal{A}$, cos $\iota$, and $\lambda$
independently and reduce the $7
\times 7$ matrix to a $3 \times 3$ matrix, which is easier to
interpret\footnote{Another reason to consider a smaller matrix is
  that it is difficult to stabilise $\mathcal{C}$ for low
  inclinations $\iota \leq 10^{\circ}$. We find that the optimal
  choice of $d\theta_i$ for which the parameter uncertainties
  $\sigma_i$ are stable varies as a function of inclination. For low
  values of $\iota$, the optimal $d\theta_i$ is orders of
  magnitude larger than for high values of $\iota$. Note that the FIM
  (as opposed to $\mathcal{C}$) is stable for the same choice of
  $d\theta_i$. In addition, the FIM choice of $d\theta_i$ that is made for
  the inclination affects only matrix elements 
  that contain terms related to $\iota$. However, this influence spreads to
  all matrix elements of $\mathcal{C}(=\Gamma^{-1})$, because of the
  matrix inversion.}. 
In Figure~\ref{fig:AMCVn_errs_func_i_3X3}, we show the S/N, 
(relative) parameter uncertainties and the normalised correlation for these three 
parameters as a function of inclination. It is obvious that for a
source at a given distance, the S/N for an edge-on binary is lower than that 
for a face-on binary owing to the less favourable orientation of the latter. Thus, the
S/N is a function of $\cos \iota$ and has a minimum at 
$\iota = 90^{\circ}$ $(\cos\iota = 0)$. In general, we expect the parameter
uncertainties to vary accordingly, i.e., a higher S/N should correspond to
a higher precision and \textit{vice-versa}. The longitude 
  follows this behaviour in the bottom-left panel. However, for
$\mathcal{A}$ and $\cos \iota$ this is not the
case as the detector cannot accurately determine the inclination in
the range of $0.7\lesssim |\cos\iota| \leq 1$, as shown in the
  top-right and middle-left panels. For $\iota
\le 10^{\circ}$ and $\iota \ge 170^{\circ}$ in particular, the uncertainties in
$\iota$ and $\mathcal{A}$ become very large. However, as shown in the
middle-right panel, the correlations are maximal for these almost
face-on binaries, thus by constraining one of the parameters from
EM (or other) data, we constrain the range of values that the other
parameter can take. This method cannot be applied to edge-on binaries
as the correlation is very weak in these cases. If we take a value of
$d\iota$ for which the computation of the $3 \times 3 $
variance-covariance matrix is numerically stable for all $\iota$ and
use it to produce the full $7 \times 7$ matrix, the parameter 
uncertainties behave as shown in
Figure~\ref{fig:AMCVn_errs_func_i_7X7}.  The uncertainties
$\sigma_{\phi_0}$ and $\sigma_{\psi}$ behave similarly to
$\sigma_{\mathcal{A}}$ and $\sigma_{\iota}$ since $\phi_0$, and $\psi$ are
also degenerate for nearly face-on systems. In contrast, the 
uncertainties $ \sigma_{ f}, \sigma_{\lambda}$, and $\sigma_{\sin
  \beta}$ behave as may be expected from the S/N since they are not strongly
tied to any of the four parameters with strong correlations. 
\begin{figure*}
\centering
\includegraphics[width=17cm]{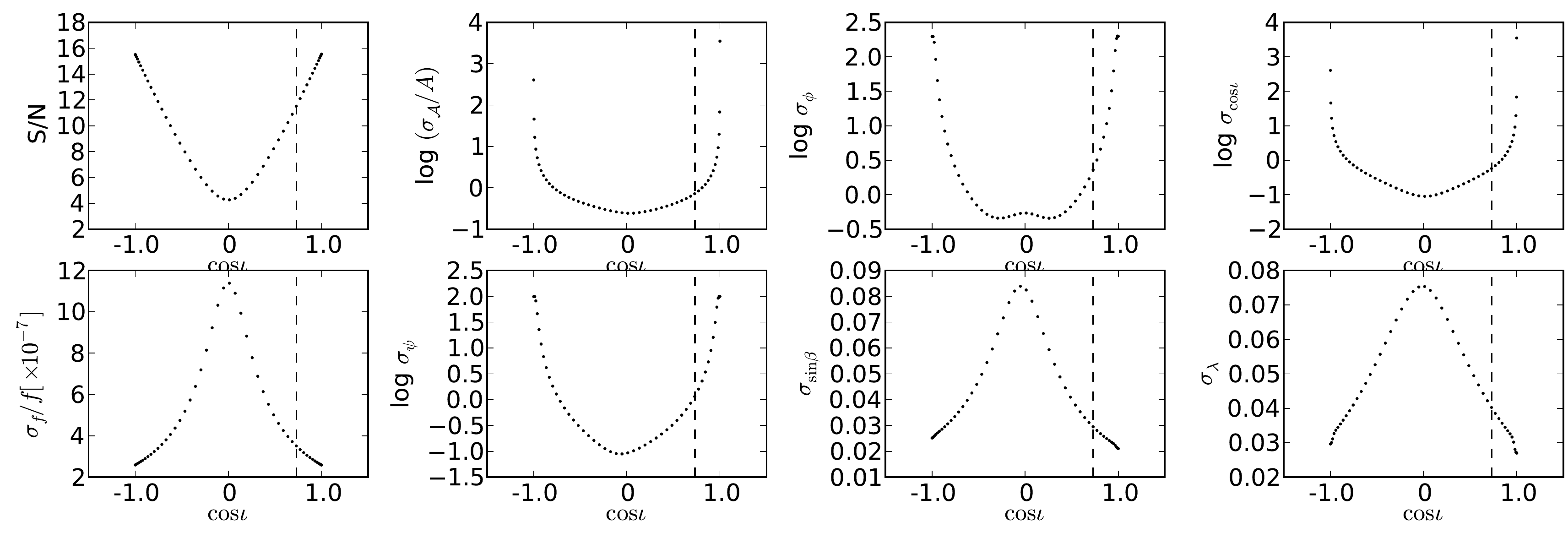}
\caption{Signal-to-noise ratio and (relative) parameter uncertainties for AM CVn  as a
    function of $\cos \iota$. The general behaviour for
    $\sigma_{\mathcal{A}}$, $\sigma_{\mathrm{cos}\iota}$ and
    $\sigma_{\lambda}$ is consistent with
  Figure~\ref{fig:AMCVn_errs_func_i_3X3} above. The uncertainties in 
  $\sigma_{\mathrm{sin}\beta}, \sigma_{\lambda},$ and  $\sigma_{ f}$ behave
  as may be expected from the S/N,  where parameters are more
  accurately determined for higher S/N. }
\label{fig:AMCVn_errs_func_i_7X7}
\end{figure*}

\subsection{The correlation between $\phi_0$ and $\psi$}
The orientation parameters $\phi_0$ and $\psi$ are also highly
anti-correlated with $c_{\phi_0 \psi} \approx - 0.99$ for AM~CVn and
HM~Cnc. This can also be explained by examining the geometry of the
physical binary. The orbital phase $\phi$ is measured in the plane of
the binary (perpendicular to the binary's angular momentum vector,
$\hat{L}$), while $\psi$ is measured in the plane
perpendicular to the line of sight, $\hat{n}$. Both AM~CVn and HM~Cnc
have an inclination in the range where the system appears to be
face-on to the detector (see Section 3.1). Thus, $\phi_0$ and $\psi$
are practically degenerate for these binaries. In contrast, for J0651
$\hat{L}$ and $\hat{n}$ are almost perpendicular, and the effects of
$\phi_0$ and $\psi$ can be more easily distinguished by \emph{eLISA}. This
correlation is not very noteworthy, because the parameters involved
are of little astrophysical interest.  

\subsection{The correlation between $f $ and $\phi_0$}
For J0651, there is a strong anti-correlation between $f$ and
$\phi_0$, with a value of $- 0.89$. This is because if the frequency of a
signal is changed by a small fraction, $\delta f$, the 
signal accumulates a phase shift towards the end of the wave. This 
signal can be de-shifted in phase to obtain another signal with a similar
match to the original, in turn making the two slightly dissimilar
signals indistinguishable. Hence, the two parameters $f $ and $\phi_0$ are
degenerate. This degeneracy is not present for AM~CVn and HM~Cnc
owing to the strong anti-correlation between $\phi_0$ and $\psi$ and
large errors in these two parameters. 

\subsection{The correlation between $f$ and $ \dot{f}$}
If we model HM~Cnc with an additional eighth parameter, $\dot{f}$, we
find a strong anti-correlation between $f$ and $\dot{f}$. This can
be explained in a very similar way to the case of $c_{f \phi_0}$
above. Consider a signal which is obtained by changing the chirp by
$\delta \dot{f}$, within its uncertainty. This will slightly change 
the frequency towards the end of the signal. By additionally changing $f$
within its uncertainty, we can obtain a very similar signal to the one
for which we changed only its chirp by $\delta \dot{f}$. Thus, the two
parameters are degenerate for a mildly chirping binary. This could be a very
useful correlation for constraining the $\dot{f}$ of the binary by using an EM
constraint on $f$. However, the GW data analysis already provides
a relative accuracy of $\sim 10^{-6}$ for the frequency, which is hard to
improve with EM observations. 

After providing a potential measurement of $\dot{f}$ for the J0651 source (M. Kilic,
private communication), the correlation between $f$ and
$\dot{f}$ could be used to test one of the predictions of General 
Relativity in Eq.\,\ref{eq:fdot}, which holds for a binary
that evolves only because of GW radiation loss. Given that the individual masses and
their $f$ are measured, Eq.\,\ref{eq:fdot} can be used to test the prediction of $\dot{f}$.

\section{Discussion}
The S/N's of the data for the verification binaries listed in
Table~\ref{tab:verf_bin} are relatively low. In particular,  J0651 has
S/N $\sim 10$. This may not satisfy S/N $\gg 1$, one 
of the criteria under which FIM studies are valid. To exclude the effects
of S/N, we place the verification binaries at a closer distance so
that they all have S/N of $\sim 100$. We find that all of the parameter
uncertainties decrease as expected, except for the uncertainty in
$\mathcal{A}$. However, the strong correlations we
found for $c_{\mathcal{A}\iota}$, $c_{\phi_0 \psi}$,  $c_{f \phi_0}$
and $c_{f \dot{f}}$ listed above in the matrices remain the same. 
Furthermore, these correlations do not change when we consider the
\textit{six-link} configuration of the classic \textit{LISA} with a five
million km baseline and slightly different instrumental noise. From
the \textit{six-link}  interferometer, three independent data streams can be 
synthesised, of which \emph{A, E,} and \emph{T} is one of the combinations. Naturally,
this means that the S/N of the source will be higher, and that there
may be additional information from these data streams. However, we
found that when we consider these optimal TDIs, the strong correlations we
found are unaffected. Additionally, the weak correlations
  between the rest of the parameters are also unaffected.

It is well-known that in the low-frequency regime there are
  many overlapping sources, which can significantly degrade the
  parameter uncertainty. Crowder \& Cornish (2004) performed a 
thorough FIM analysis and show that the degradation of parameter uncertainties
depends on the frequency, the number of binaries per frequency
bin, and the observation time. However, the degradation is uniform for all the seven
parameters, which means that each parameter is affected in
the same way by the overlapping sources. Hence, the presence of
overlapping sources affects the \textit{uncertainties} in the parameters, but
not the normalised \textit{correlation} between the
parameters. In addition the potential biases from imperfect
subtraction of resolvable sources (Section 2.3) can affect the
results, although this mostly affects the S/N degradation of the source
and thus not the normalised correlations.

Our study should be used to implement EM
  priors in an MCMC data
  analysis based on a Bayesian framework. While the latter leads to more
  accurate results, the FIM tools provides useful predictions 
  \citep{2008PhRvD..77d2001V}. In the case of Galactic
  binaries, it has been shown that the one-dimensional marginalised posterior PDFs
  are well-matched by the FIM predictions when the data analysis of  a
  single source is considered \citep{2005PhRvD..72d3005C}. 
 
In a forthcoming paper, we will discuss the dependence of
the aforementioned correlations
and accuracies on the sky position and polarisation angle. Our test
calculations show that the strong correlations $c_{\mathcal{A}\iota}$,
$c_{\phi_0\psi}$, $c_{f\phi_0}$, and $c_{f\dot{f}}$ found above do not
vary as a function of these parameters.  

Finally, we observe that the uncertainty in the inclination of a
binary indicates whether the system is eclipsing. In Section 3.1,
we found that for J0651, the uncertainties in $\iota, \phi_0$, and
$\psi$ are small compared to those of HM~Cnc, even though its
S/N is a factor of three lower than that of HM~Cnc. An uncertainty of $\sigma_{\cos
  \iota} = 0.043$ translates to $\sigma_{\iota} \sim 2.5^{\circ}$. 
Hence, for J0651-like systems with S/N's of $\gtrsim 10$, the small uncertainty 
inclination inferred from a GW analysis may be used to find candidate eclipsers.  

\section{Conclusion}
We have performed Fisher-matrix studies to investigate whether there are
correlations between the parameters that characterise Galactic binaries
with short periods, between about six minutes and a few hours, which lie in the \emph{eLISA}
frequency band, $10^{-4} $ Hz $- 1 $ Hz. We focused on three verification
binaries, AM~CVn, SDSS~J0651+2844, and the mildly chirping HM~Cnc. Our
main findings are: 
\begin{enumerate}
\item There are strong correlations between the strain amplitude and
  inclination, and between the phase and polarisation angle for AM~CVn and
  HM~Cnc. For the latter, there is an additional strong anti-correlation
  between $f$ and $\dot{f}$ when it is modelled as a mildly chirping
  source.
\item The remaining parameters are not or only weakly correlated
  for the binaries considered. 
\item The correlation between strain amplitude and inclination can be very
  useful in constraining the amplitude of the binary, which is a
  function of the two masses, the distance to the binary, and its
  frequency. Since the frequency is very accurately determined (of
  the order of $10^{-9} \:$Hz) by the GW data analysis, a combination of
  the 
  masses and the distance can be constrained by using an EM constraint
  on the inclination.
\item These correlations depend strongly on the inclination of the
  system: approximately face-on ($\iota \lesssim 45^{\circ}, \; \iota
  \gtrsim 135^{\circ}$)
  binaries have strongly correlated parameters, whereas
  for edge-on binaries, the correlations become very weak.
\item For binaries with very low inclinations ($\iota < 10^{\circ}$
  and $\iota > 170^{\circ}$), the uncertainties in
  amplitude, inclination, phase, and polarisation angle derived from
  GW data are very large. 
\item We have found that the influence of S/N on the correlations is 
  not significant; placing the sources at shorter distances to give them
  higher S/N yields the same normalised correlations.
\item The strong correlation between amplitude and inclination is the
  most useful correlation found, since inclinations are typically
  measured to a higher accuracy by EM observations than with GW
  measurements alone. Hence, this correlation can be
  used to constrain the amplitude by a factor of six for systems with
  orientations and S/N's similar to those of AM~CVn.  
\item We have found that some correlations also depend on sky position and
  polarisation angle, which we will discuss in a forthcoming paper.
\item Even for signals with an S/N of $\sim 10$, we have been able to
  reliably determine the inclination when the system is edge-on. This
  will enable efficient searches for eclipsing systems with EM instruments.
\end{enumerate}

\bibliographystyle{aa} 
\bibliography{literature_data_analysis,literature_binary_science}

\begin{acknowledgements}
This work was supported by funding from FOM. We are very grateful to
Michele Vallisneri for providing support with the \textit{Synthetic LISA} and
\textit{Lisasolve} softwares. We thank the anonymous referee for
the helpful suggestions to improve the paper. 
\end{acknowledgements}

\begin{appendix}
\section{Stabilisation of the variance-covariance matrix}
Here we give an example of the Stabilisation of the variance-covariance
matrix for AM~CVn. The $d\theta{_i}$ are varied for each parameter
over seven orders of magnitude as described in Section 2.4. In this
example, $d\theta_{\mathcal{A}}$, $d\theta_{f}$, $d\theta_{\iota}$,
$d\theta_{\psi}$ are varied from $(10^{-4}, ... \: , 10^{2}) \times
\sqrt{\epsilon} \times \theta_i$, where, $\theta_i$ are the true values
for AM~CVn. The values of $d\theta_{\phi}$, $d\theta_{\beta}$,
$d\theta_{\lambda}$ are varied from $10^{-6}, ... \: , 10^{0}$, more 
than for parameters above, because the 
numerical derivatives of $\phi$, $\beta$, and $\lambda$ are unstable
for lower values . The first row of the uncertainties in the
parameters correspond to the first set 
of $d\theta{_i}$, et cetera. In principle, the uncertainties 
$\sigma_i$ should level off around the third 
row. The uncertainties for all our verification binaries are
calculated in this way. 

\begin{table}[!h]
\begin{center}
\caption{Parameter uncertainties, $\sigma_i$, in a range of $d
  \theta_i$ for the case of AM CVn, with S/N $\sim 11$ for
  $T_{\mathrm{obs}} = 2$ years. The values start to
  level off from the third to fifth row.}
\label{tab:stabilise}
      \begin{tabular}{c c c c c c c }
    \hline  \hline
$\sigma_{\mathcal{A}}$ & $\sigma_{\phi}$ & $\sigma_{\mathrm{cos}\iota}$ &$\sigma_{f}$& $\sigma_{\psi}$  & $\sigma_{\mathrm{sin}\beta}$ & $\sigma_{\lambda}$ \\ \hline
$1.417\times 10^{-23}$& $0.317$& $0.072$& $6.517\times 10^{-10}$& $0.128$& $0.027$& $0.029$\\ \hline
$7.934\times 10^{-23}$& $1.693$& $0.426$& $6.533\times 10^{-10}$& $0.846$& $0.028$& $0.034$\\ \hline
$1.037\times 10^{-22}$& $2.234$& $0.555$& $6.541\times 10^{-10}$& $1.120$& $0.028$& $0.038$\\ \hline
$1.041\times 10^{-22}$& $2.242$& $0.557$& $6.541\times 10^{-10}$& $1.124$& $0.028$& $0.039$\\ \hline
$1.041\times 10^{-22}$& $2.242$& $0.557$& $6.541\times 10^{-10}$& $1.124$& $0.028$& $0.039$\\ \hline
$1.035\times 10^{-22}$& $2.250$& $0.554$& $6.556\times 10^{-10}$& $1.126$& $0.029$& $0.04$\\ \hline
$8.756\times 10^{-23}$& $2.301$& $0.473$& $8.259\times 10^{-10}$& $0.942$& $0.113$& $0.122$\\ \hline
  \end{tabular}
\end{center}
\end{table}

\section{Sampling}
\begin{figure}[!h]
\centering
\includegraphics[width=\columnwidth]{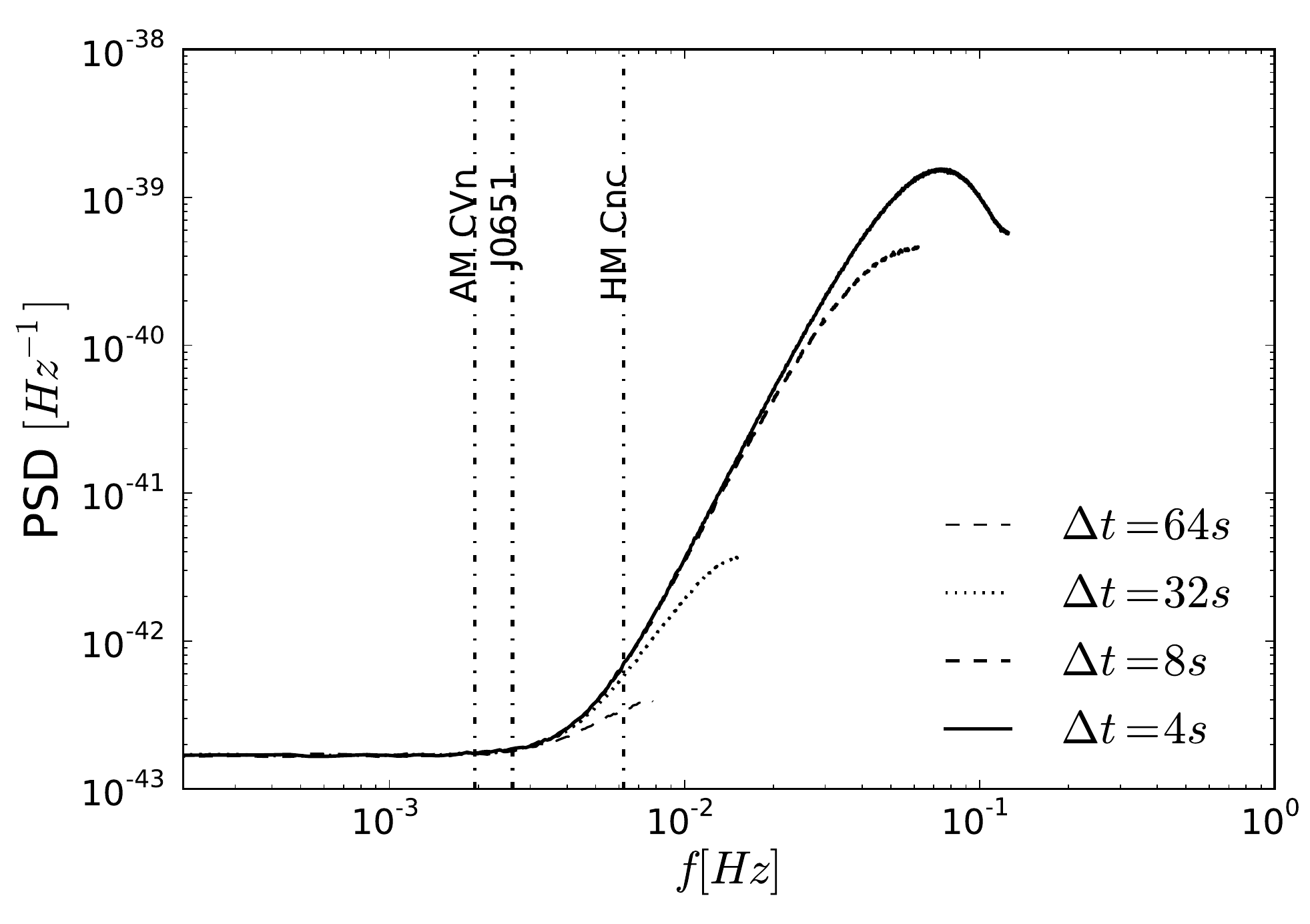}
\caption{Average PSDs of instrumental noise for \emph{eLISA} with varying sampling
  time, $\Delta t$. For lower-frequency sources such as AM~CVn and J0651,
a sampling as fine as $64$s can be used. However, for HM~Cnc, whose frequency
is higher, a high $\Delta t$ would overestimate its S/N, 
since the level of the noise is too low for the longest sampling times.  }
\label{fig:inst_noises} 
\end{figure}

In Fig~\ref{fig:inst_noises}, we show the effect of the sampling time
$\Delta t$ on the noise PSD of the instrumental noise, which is the average spectrum of
the TDI $X$ observable. At $f < 5\times10^{-3}$ Hz, the sampling time
does not affect the level of the PSD. Above these frequencies,
the PSD with a relatively long sampling time of $64$s underestimates
its true level.

\end{appendix}

\end{document}